\documentclass[acmtog, nonacm]{acmart}
\usepackage{bm}
\usepackage{multirow}
\usepackage{soul}
\usepackage{amsmath}
\usepackage{pifont}
\usepackage{subcaption}
\usepackage{makecell}
\usepackage[capitalise]{cleveref}

\AtBeginDocument{%
  }

\newcommand{\ours}{DC-VSR}

\newcommand{\ssi}{Spatial Attention Propagation}
\newcommand{\tfi}{Temporal Attention Propagation}
\newcommand{\drg}{Detail-Suppression Self-Attention Guidance}

\newcommand{\ssiabb}{SAP}
\newcommand{\tfiabb}{TAP}
\newcommand{\drgabb}{DSSAG}

\newcommand{\etal}{{\textit{et al.}}}

\begin{document}

\title{\ours{}: Spatially and Temporally Consistent Video Super-Resolution with Video Diffusion Prior}

\author{Janghyeok Han}
\authornote{Both authors contributed equally to this research.}
\affiliation{
  \institution{POSTECH}
  \country{Republic of Korea}
}
\email{hjh9902@postech.ac.kr}

\author{Gyujin Sim}
\authornotemark[1]
\affiliation{
  \institution{POSTECH}
  \country{Republic of Korea}
}
\email{sgj0402@postech.ac.kr}

\author{Geonung Kim}
\affiliation{
  \institution{POSTECH}
  \country{Republic of Korea}
}
\email{k2woong92@postech.ac.kr}

\author{Hyun-Seung Lee}
\affiliation{
  \institution{Visual Display Business, Samsung Electronics}
  \country{Republic of Korea}
}
\email{hyuns.lee@samsung.com}

\author{Kyuha Choi}
\affiliation{
  \institution{Visual Display Business, Samsung Electronics}
  \country{Republic of Korea}
}
\email{kyuha75.choi@samsung.com}

\author{Youngseok Han}
\affiliation{
  \institution{Visual Display Business, Samsung Electronics}
  \country{Republic of Korea}
}
\email{yseok.han@samsung.com}

\author{Sunghyun Cho}
\affiliation{
  \institution{POSTECH}
  \country{Republic of Korea}
}
\email{s.cho@postech.ac.kr}

\begin{abstract}
    Video super-resolution (VSR) aims to reconstruct a high-resolution (HR) video from a low-resolution (LR) counterpart. Achieving successful VSR requires producing realistic HR details and ensuring both spatial and temporal consistency. To restore realistic details, diffusion-based VSR approaches have recently been proposed. However, the inherent randomness of diffusion, combined with their tile-based approach, often leads to spatio-temporal inconsistencies. In this paper, we propose \ours{}, a novel VSR approach to produce spatially and temporally consistent VSR results with realistic textures. To achieve spatial and temporal consistency, \ours{} adopts a novel Spatial Attention Propagation (SAP) scheme and a Temporal Attention Propagation (TAP) scheme that propagate information across spatio-temporal tiles based on the self-attention mechanism. To enhance high-frequency details, we also introduce Detail-Suppression Self-Attention Guidance (DSSAG), a novel diffusion guidance scheme. Comprehensive experiments demonstrate that \ours{} achieves spatially and temporally consistent, high-quality VSR results, outperforming previous approaches. The project page is available at \href{https://daramgc.github.io/docs/Publications/dc-vsr}{\textcolor{blue}{https://daramgc.github.io/docs/Publications/dc-vsr}}
\end{abstract}



\begin{teaserfigure}
  \centering
  \includegraphics[width=0.95\textwidth]{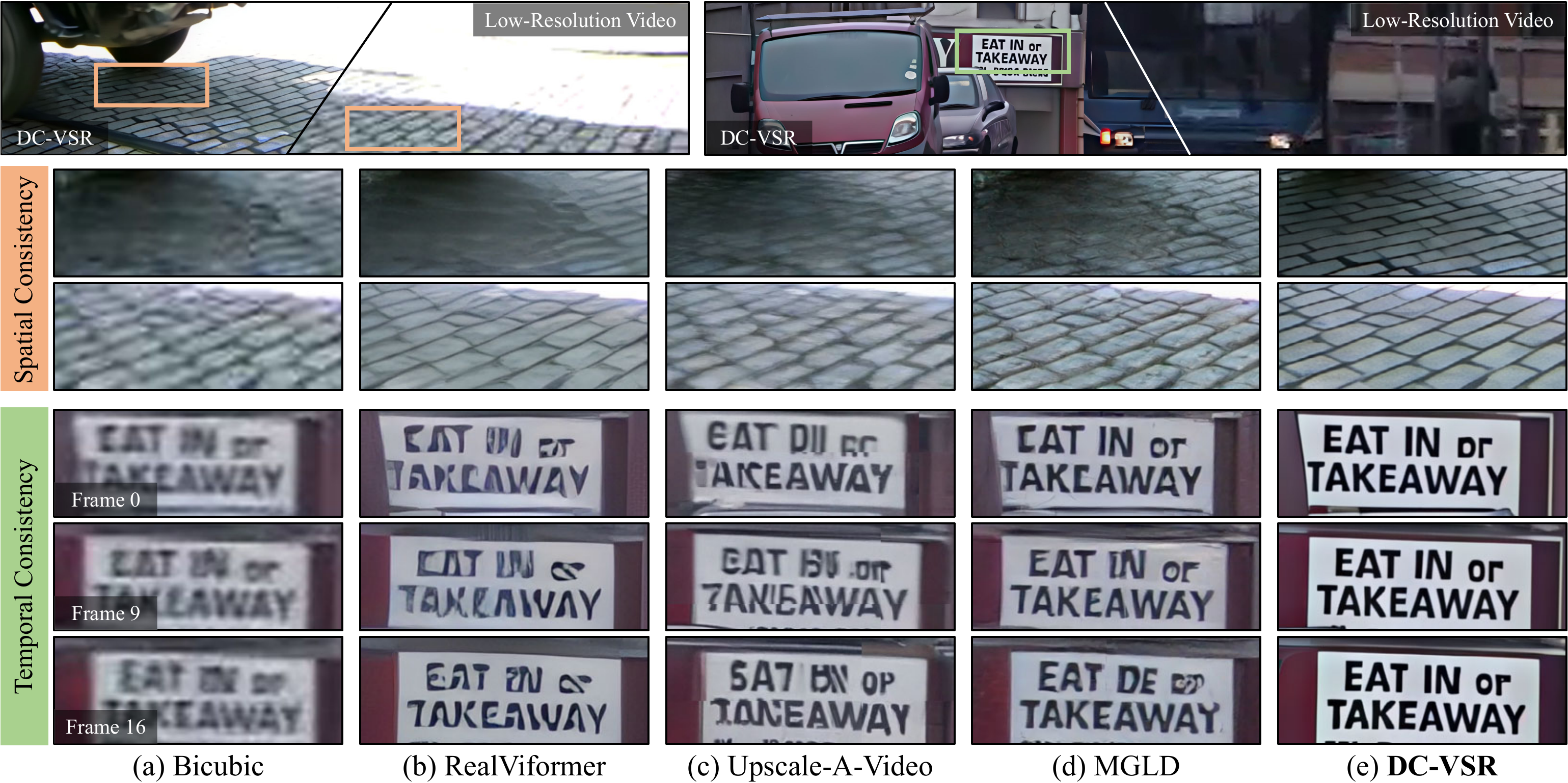}
  \vspace{-2mm}
  \caption{Real-world video super-resolution results of our proposed methods and state-of-the-art models: RealViformer~\cite{realviformer}, Upscale-A-Video~\cite{zhou2024upscaleavideo}, MGLD~\cite{yang2023mgldvsr}. \ours{} shows outstanding super-resolution performance, in respect of not only quality but also spatial and temporal consistency. The input videos are from the VideoLQ dataset~\cite{chan2022investigating} (samples 037 and 038).
  }
  \Description{teaser}
  \label{fig:teaser}
\end{teaserfigure}

\maketitle

\section{Introduction}
\label{sec:intro}

Video super-resolution (VSR) is a task to restore a high-resolution (HR) video from a low-resolution (LR) counterpart, which has a vast array of applications, such as enhancing old footage and improving streaming video quality on limited bandwidths.
However, VSR is particularly challenging due to its severely ill-posed nature of the problem, primarily because of the missing high-frequency information in LR videos caused during the sampling process.
Furthermore, real-world videos face a myriad of unknown degradation factors beyond just sampling issues, including noise, compression artifacts, and various other distortions, making the task even more ill-posed.

To achieve successful VSR, it is essential to generate realistic HR details while maintaining both spatial and temporal consistency.
To this end, over the past few decades, numerous approaches have been proposed, including traditional methods such as interpolation techniques, model-based optimization approaches, and recent neural network-based ones~\cite{realviformer}.
Nevertheless, most of them struggle to produce realistic HR details due to a lack of effective priors that model high-frequency details as shown in \cref{fig:teaser}(b).

Recently, diffusion models that provide powerful generative priors for natural images have been exploited for VSR to achieve high-quality results with realistic textures.
Yang \etal~\shortcite{yang2023mgldvsr} and Zhou \etal~\shortcite{zhou2024upscaleavideo} utilize image diffusion models and successfully restore detail-rich HR videos.
However, adopting image diffusion models poses challenges in maintaining spatial and temporal consistency due to the design of image diffusion models, which typically target single images with limited spatial sizes due to significant memory usage.
To achieve temporally-consistent VSR using image diffusion models, these approaches adopt additional temporal layers, temporal propagation with motion compensation, and overlapping of temporal windows.
Furthermore, to handle large video frames, frames are split into overlapping tiles, processed individually, and then merged.
Despite these efforts, resulting videos often suffer from spatial and temporal inconsistencies due to the inherent randomness of image diffusion models and their tile-based approach (\cref{fig:teaser}(c) and (d)).

In this paper, we propose \emph{\ours{} (Diffusion-based Consistent VSR)}, a novel VSR approach that, for the first time, leverages a video diffusion prior.
Our approach produces spatially and temporally consistent results with realistic textures, given an LR video of arbitrary length and spatial size (\cref{fig:teaser}(e)).
We employ Stable Video Diffusion (SVD)~\cite{svd} as a video diffusion prior, which provides powerful generative capabilities for restoring high-quality, temporally-consistent details.
However, exploiting an existing video diffusion model for consistent VSR over long video clips with large frames is challenging, as existing models are designed to synthesize a limited number of small  frames, similar to image diffusion models.

To achieve spatial and temporal consistency for a long video with large frames, \ours{} introduces a novel Spatial Attention Propagation (SAP) scheme and Temporal Attention Propagation (TAP) scheme.
Specifically, \ours{} decomposes an input LR video into multiple spatio-temporal tiles, and processes them separately.
To achieve spatial consistency across tiles, SAP introduces a subsampled feature map representing the entire area of a video frame and uses it to process tiles at different spatial locations.
On the other hand, TAP enhances temporal consistency across tiles by propagating information between temporally consecutive tiles.
Both schemes are realized by extending the self-attention layers of a video diffusion model, enabling information on HR details to be effectively propagated across tiles without losing the generative capability of a pretrained diffusion model.

Additionally, we propose Detail-Suppression Self-Attention Guidance (DSSAG), a novel diffusion guidance scheme to improve high-frequency details in synthesized HR video frames.
Similar to Self-Attention Guidance (SAG)~\cite{sag} and Perturbed Attention Guidance (PAG)~\cite{pag}, DSSAG guides the diffusion process by amplifying high-frequency details in the latent representation.
However, unlike the previous methods, DSSAG provides more flexible control over the guidance scale, and can seamlessly integrate with classifier-free guidance (CFG) for high-quality synthesis without incurring additional computational overhead.

We demonstrate the effectiveness of our approach on real-world VSR tasks, where input LR videos contain various unknown degradations. Our experimental results show that \ours{} achieves spatially and temporally consistent, high-quality VSR results, outperforming previous approaches. Our contributions are summarized as follows:
\begin{itemize}
    \item We introduce DC-VSR, a novel VSR approach based on a video diffusion prior, which produces spatially and temporally consistent results with realistic textures. Our approach is the first to exploit a video diffusion prior in VSR.
    \item We propose SAP that injects subsampled features representing the entire area of a video frame to different tiles, ensuring spatial consistency.
    \item We propose TAP for sharing information across temporally distant frames, achieving temporal consistency.
    \item We introduce DSSAG to enhance the quality of synthesized video frames without additional computational overhead.
\end{itemize}

\begin{figure*}[t]
    \centering
    \includegraphics[width=0.90\linewidth]{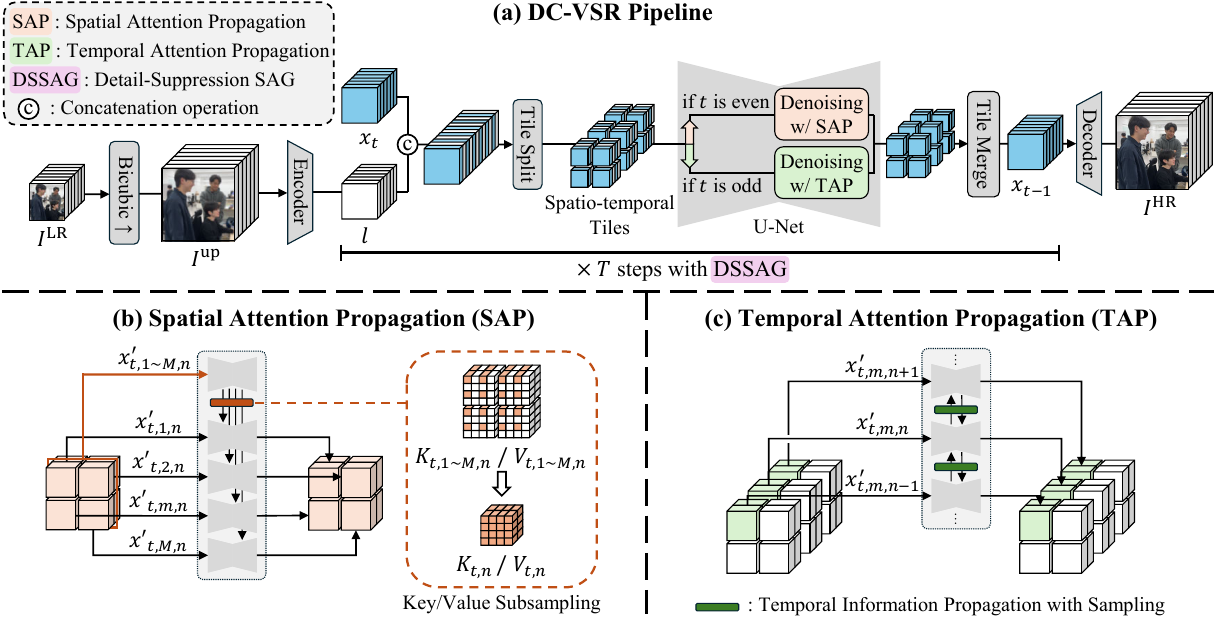}
    \vspace{-2mm}
    \caption{Overall pipeline of proposed {\ours}. An encoded low-resolution video latent is concatenated to the current noisy latent $x_t$, and it undergoes alternating denoising processes using both {\ssi} (\ssiabb) and {\tfi} (\tfiabb). At this stage, the noisy latent is split and merged before and after the denoising process, respectively. 
    After each denoising step, {\drg} ({\drgabb}) is applied to enhance the quality of the image further.
    }
    \vspace{-2mm}
    \Description{pipeline}
    \label{fig:main}
\end{figure*}

\section{Related Work}
\paragraph{Generative Prior for Video Super-Resolution}
VSR methods can broadly be classified into two main categories based on the use of generative prior. VSR models without generative prior \cite{Youk_2024_CVPR, chan2021basicvsr, chan2022basicvsrpp, chan2022investigating, fastrealVSR, realviformer} are typically trained using reconstruction losses, such as the mean squared error (MSE), augmented with additional techniques designed to enhance temporal consistency. Although they restore HR videos closely matching the input LR videos, they fail to generate complex details, producing blurry results due to the significant ill-posed nature of the problem. To achieve VSR with rich details, generative prior-based VSR methods \cite{rota2024stablevsr, xu2024videogigagan, chen2024sateco, zhou2024upscaleavideo, yang2023mgldvsr} have recently emerged. These methods fine-tune GANs \cite{gan} or diffusion-based image generation models by modifying them to take LR inputs as conditions and incorporating additional modules to consider temporal consistency. However, maintaining consistency across frames with detailed textures is still challenging \cite{blattmann2023stablevideodiffusionscaling,blattmann2023videoldm}, leading to temporal flickering artifacts.

\paragraph{Image Super-Resolution}
As deep learning techniques have advanced, various image super-resolution (ISR) methods have been proposed. SRCNN~\cite{srcnn} was the first to introduce a CNN-based SR method. Subsequent studies have applied various network architectures, such as residual networks~\cite{vdsr,srgan}, dense networks~\cite{rdn,esrgan,real-esrgan}, Laplacian pyramid networks~\cite{lapsrn}, back-projection networks~\cite{dbpn}, recursive structures~\cite{drcn,drrn}, and channel attention~\cite{rcan}. While these methods have demonstrated promising SR results, they often struggle to recover high-frequency details. To address this, approaches leveraging the generative prior of pretrained generative models have been introduced exploiting GANs~\cite{glean,pulse} or Diffusion models~\cite{stablesr,supir,diffbir,pasd,seesr,ccsr,wu2024one}. Since the application of these methods to videos leads to severe flickering artifacts due to the absence of temporal consistency, VSR research has been active in addressing this issue.

\paragraph{Tile-based Generation}
Diffusion models face two main challenges in generating higher spatial or temporal resolutions: quality degradation from discrepancies between training and inference conditions, and substantial computational costs. To address these challenges, tile-based generation approaches have been proposed~\cite{multidiffusion2023,syncdiffusion2023,demofusion2024,zhou2024upscaleavideo,yang2023mgldvsr}. These methods divide the input into overlapping tiles, perform diffusion sampling on each tile individually, and subsequently merge the tiles. However, their approaches do not consider inter-tile consistency, often generating inconsistent details for the same content across split tiles in an image or distant frames~\cite{zhou2024upscaleavideo,yang2023mgldvsr}.

\section{\ours{}}
\label{sec:method}
Given an input LR video $I^\textrm{LR}$ consisting of $N$ frames $\{I^\textrm{LR}_1, \cdots, I^\textrm{LR}_N\}$, \ours{} produces an HR video $I^\textrm{HR}$.
\cref{fig:main} illustrates the overall framework of \ours{}, which is built upon the SVD framework~\cite{svd}.
To begin, \ours{} upsamples $I^{LR}$ using bicubic interpolation to match the target resolution, and obtains an upscaled video $I^\textrm{up}$.
It then embeds $I^\textrm{up}$ into the latent space using the VAE encoder~\cite{ldm}, obtaining a latent representation $l$, which consists of $[l_1, \cdots, l_N]$ stacked along the channel dimension where $l_i$ represents the latent of the $i$-th upsampled video frame $I^\textrm{up}_i$.
To generate an HR video, \ours{} initializes the latent representation $x_T$ of $I^\textrm{HR}$ as random noise, where $T$ is the number of diffusion sampling steps, and $x_T$ is a tensor with the same size as $l$.

At each diffusion sampling step $t$,
$x_t$ and $l$ are first concatenated in an interleaved manner, i.e., $[x_{t,1}, l_1, \cdots, x_{t,N}, l_N]$, where $x_{t,i}$ is the noisy latent of the $i$-th HR video frame.
The concatenated latents are then split into spatio-temporal tiles.
We refer to each tile as $x'_{t,m,n}$ where $m$ and $n$ are spatial and temporal indices.
Each tile is processed through a denoising U-Net, and the processed tiles are merged to obtain the latent representation $x_{t-1}$ at the next sampling step $t-1$.
We utilize spatio-temporal tiles of size $64\times64\times14$ in the latent space, corresponding to $512\times512\times14$ in the image space with a scaling factor of 8. In line with previous approaches~\cite{zhou2024upscaleavideo, yang2023mgldvsr}, spatially and temporally neighboring tiles overlap by 50\%.
Overlapped tiles are blended in the tile merging step in our pipeline using alpha blending with Gaussian weights.
To achieve spatial and temporal consistency, \ours{} alternatingly applies either SAP or TAP at each sampling step until $t$ reaches $0$.
Finally, $x_0$ is fed to a decoder, producing the HR video $I^\textrm{HR}$.

\ours{} employs a tile-based approach to handle lengthy videos with large frames with a video diffusion prior.
However, na\"ively splitting a video into tiles may introduce spatial and temporal inconsistencies.
In image diffusion models like latent diffusion model (LDM)~\cite{ldm}, self-attention layers of the denoising U-Nets ensure spatial consistency within an image.
Likewise, video diffusion models such as SVD~\cite{svd} leverage self-attention to achieve spatially and temporally coherent results.
Specifically, the self-attention operation is defined as:
\begin{equation}
    \text{SA}(\bm{Q},\bm{K},\bm{V}) = \text{softmax} \left( \frac{\bm{QK}^\top}{\sqrt{d}} \right) \bm{V},
    \label{eq:self_attention}
\end{equation}
where $\bm{Q}$, $\bm{K}$ and $\bm{V}$ are query, key, and value in the matrix representations, respectively, and $d$ is the attention feature dimension.
For a certain spatial and temporal position in a video, the self-attention operation calculates the correlation between the query at position and keys at other positions, and aggregates values based on these correlations. As a result, the synthesized content of any region within the video is harmonized with the rest of the video.
However, when a video is split into tiles, each tile undergoes an independent attention process, resulting in spatial and temporal inconsistencies. To address this, \ours{} extends the self-attention operations using SAP and TAP, allowing attentions to be efficiently computed across tiles.
In \cref{ssec:SSI,ssec:TFI}, we describe SAP and TAP to achieve spatial and temporal consistency across tiles.
We then explain DSSAG to further enhance VSR quality in \cref{ssec:DSSAG}.

\subsection{Spatial Attention Propagation}
\label{ssec:SSI}
To achieve spatial consistency across tiles, SAP extends self-attention operations for each tile to incorporate information from different tiles. However, due to the quadratic computational complexity of attention, na\"ive extension of self-attention operations is practically infeasible.
Instead, to avoid the quadratic increase of the computational complexity, SAP leverages subsampled features that represent the entire areas of video frames and injects them into the self-attention operations for each tile.

\cref{fig:main}(b) illustrates the SAP scheme.
At diffusion sampling step $t$ for SAP, we feed each tile $x'_{t,m,n}$ to the denoising U-Net.
Then, at each self-attention layer, we compute key/value pairs, and subsample them in a spatially uniform manner with respect to a predefined sampling rate $s_\textrm{SAP}$.
Finally, we aggregate the subsampled key/value pairs for all $m \in \{1,\cdots,M\}$, obtaining the subsampled sets of keys and values, $K_{t,n}$ and $V_{t,n}$.
Once $K_{t,n}$ and $V_{t,n}$ are obtained, we inject them into the self-attention operation of each tile.
Specifically, let us denote the query, key, and value sets computed from tile $x'_{t,m,n}$ as $Q_{t,m,n}$, $K_{t,m,n}$, and $V_{t,m,n}$, respectively.
We construct new sets of keys and values $\hat{K}_{t,m,n}$ and $\hat{V}_{t,m,n}$ by merging $K_{t,m,n}$ and $K_{t,n}$, and $V_{t,m,n}$ and $V_{t,n}$, respectively.
Finally, we perform the self-attention operation using the extended keys and values, i.e.,
\begin{equation}
    \text{SA}(\bm{Q}_{t,m,n},\bm{\hat{K}}_{t,m,n},\bm{\hat{V}}_{t,m,n}) = \text{softmax} \left( \frac{\bm{Q}_{t,m,n}\bm{\hat{K}}_{t,m,n}^\top}{\sqrt{d}} \right) \bm{\hat{V}}_{t,m,n},
    \label{eq:self_attention_SSI}
\end{equation}
where $\bm{Q}_{t,m,n}$, $\bm{\hat{K}}_{t,m,n}$, and $\bm{\hat{V}}_{t,m,n}$ are matrix representations of $Q_{t,m,n}$, $\hat{K}_{t,m,n}$ and $\hat{V}_{t,m,n}$, respectively.
We apply the SAP scheme specifically to the first two and last two spatial self-attention layers, as these layers play a crucial role in capturing and synthesizing HR details.

\subsection{Temporal Attention Propagation}
\label{ssec:TFI}

\cref{fig:main}(c) illustrates the TAP scheme for cross-tile temporal consistency.
TAP bidirectionally propagates information from a tile to its neighbor. Specifically, at each diffusion sampling step for TAP, the propagation is performed in either the forward or backward direction.
Without loss of generality, we describe the TAP scheme in the forward direction in the following.

At diffusion sampling step $t$, we process each tile $x'_{t,m,n-1}$ using the denoising U-Net, and extract keys and values from the self-attention layers.
We then sample a pair of subsets from the extracted keys and values, which we denote $K'_{t,m,n-1}$ and $V'_{t,m,n-1}$, respectively.
The subsampled sets $K'_{t,m,n-1}$ and $V'_{t,m,n-1}$ are then injected to the self-attention operation for the temporally subsequent tile $x'_{t,m,n}$.
Specifically, at each self-attention layer for $x'_{t,m,n}$, we construct new sets of keys and values $\hat{K}_{t,m,n}$ and $\hat{V}_{t,m,n}$ by merging $K_{t,m,n}$ and $K'_{t,m,n-1}$, and $V_{t,m,n}$ and $V'_{t,m,n-1}$, respectively.
We perform the self-attention operations using the extended keys and values using \cref{eq:self_attention_SSI}.
Similar to SAP, we apply the TAP scheme to the first two and last two spatial self-attention layers.

To sample $K'_{t,m,n-1}$ and $V'_{t,m,n-1}$, we select $L$ frames whose keys have the largest standard deviations from tile $x'_{t,m,n-1}$. This selection is based on the observation that frames with more varied and sharp details produce distinct keys, leading to larger standard deviations. In our experiments, we set $L=4$. We then use the keys and values from these samples as $K'_{t,m,n-1}$ and $V'_{t,m,n-1}$. For a detailed analysis of this sampling strategy, refer to the supplementary material.

\subsection{Detail-Suppression Self-Attention Guidance}
\label{ssec:DSSAG}

To improve the quality of VSR, \ours{} adopts DSSAG.
In this subsection, we first briefly review the previous guidance approaches: CFG~\cite{cfg}, SAG~\cite{sag}, and PAG~\cite{pag}.
We then describe DSSAG in detail.

\paragraph{CFG}
To improve sampling quality, CFG~\cite{cfg} utilizes both a conditional noise and an unconditional noise for denoising at each sampling step. Specifically, CFG is defined as:
\begin{equation}
    \epsilon_\textrm{CFG}(x_t)=\epsilon_\theta(x_t)+(1+s)(\epsilon_\theta(x_t,c)-\epsilon_\theta(x_t)),
    \label{eq:cfg}
\end{equation}
where $x_t$ is a latent of a synthesized image at diffusion sampling step $t$, $\epsilon_\theta$ is a denoising U-Net, which is parameterized by $\theta$, $s$ is the CFG scale parameter, and $c$ is the class condition.
\cref{eq:cfg} emphasizes the class-related components in the latent, resulting in the final synthesized image better reflecting the class condition $c$.

\paragraph{SAG}
Both SAG~\cite{sag} and PAG~\cite{pag} improve high-frequency details in synthesized images by introducing perturbation to the high-frequency details in the estimation of the unconditional noise. Specifically, a generalized form of the diffusion guidance can be defined as:
\begin{equation}
    \epsilon_\textrm{DG}(x_t)=\epsilon_\theta(\hat{x}_t)+(1+s)(\epsilon_\theta(\hat{x}_t, h_t)-\epsilon_\theta(\hat{x}_t)),
    \label{eq:diffusion_guidance}
\end{equation}
where $h_t$ is a condition, and $\hat{x}_t$ is a perturbed sample that lacks $h_t$.
Based on this generalized form, SAG is defined as:
\begin{equation}
    \epsilon_\textrm{SAG}(x_t)=\epsilon_\theta(b(x_t))+(1+s)(\epsilon_\theta(x_t)-\epsilon_\theta(b(x_t))),
    \label{eq:SAG}
\end{equation}
where $b$ is a blurring operation that detects local regions with high-frequency details using self-attention scores, and blurs the detected regions, while keeping the noise in $x_t$ intact.
The missing high-frequency details in $b(x_t)$ corresponds to $h_t$ in \cref{eq:diffusion_guidance}, i.e., $h_t = x_t - b(x_t)$.
\cref{eq:SAG} amplifies high-frequency details synthesized by the conditional model, eventually leading to synthesis results with higher-quality details.
SAG applies blurring only to regions with high-frequency details to keep image structure intact, as blurring the entire image may destroy image structures, causing synthesis results with inaccurate image structures. 

\paragraph{PAG}
PAG proposes a simpler approach, which is defined as:
\begin{equation}
    \epsilon_\textrm{PAG}(x_t)=\epsilon^\text{PAG}_\theta(x_t)+(1+s)(\epsilon_\theta(x_t)-\epsilon_\theta^\text{PAG}(x_t)),
    \label{eq:PAG}
\end{equation}
where $\epsilon^\text{PAG}_\theta(x_t)$ estimates noise from a perturbed version of $x_t$.
To achieve this, PAG replaces the self-attention score matrix with an identity matrix in the self-attention layers in $\epsilon^\text{PAG}_\theta(x_t)$, i.e., it replaces the self-attention operations with $SA_\text{perturb}(\bm{Q},\bm{K},\bm{V})=\bm{V}$.
As a result, $\epsilon^\text{PAG}_\theta(x_t)$ does not leverage spatially distant information for noise estimation, estimating less accurate noise from $x_t$, which is analogous to noise estimation from a perturbed version of $x_t$.

SAG and PAG noticeably improve image synthesis quality, especially when combined with CFG. However, integrating them with CFG incurs substantial computational costs. While using $\epsilon_\theta(x_t,c)$ instead of $\epsilon_\theta(x_t)$ in \cref{eq:SAG,eq:PAG} allows this combination, it requires running the denoising U-Net three times for SAG, as it needs additional running of the denoising U-Net to detect high-frequency regions for the blurring operation.
For PAG, the fixed level of perturbation complicates balancing the effects of CFG and PAG when combined. Consequently, PAG and CFG are typically applied separately, which also necessitates three runs of the denoising U-Net.

\begin{table*}[t]
    \caption{Quantitative comparisons of different VSR (x4) models on synthetic datasets (REDS4 and UDM10) and real datasets (VideoLQ). ISR, VSR and GP are image super-resolution tasks, video super-resolution tasks and generative prior. The best and second-best scores are marked in {\normalfont \textbf{bold} and \underline{underline}}. $\text{\normalfont DOVER}^*$ and $\text{\normalfont tLP}^*$ denote {\normalfont DOVER ($\times$100)} and {\normalfont tLP ($\times$100)} respectively.}
    \vspace{-2mm}
    \label{table:1}
    \scalebox{0.85}{
    \begin{tabular}{c|c|c|c|c|c|c|c|c|c|c}
    \Xhline{2\arrayrulewidth}
    \multirow{2}{*}{Datasets}&\multirow{2}{*}{Metrics}& \multicolumn{2}{c|}{ISR w/o GP}         &\multicolumn{2}{c|}{ISR w/ GP}     &\multicolumn{2}{c|}{VSR w/o GP}        &\multicolumn{3}{c}{VSR w/  GP}               \\ \cline{3-11} 
                             &                        & Bicubic             & SwinIR    & Real-ESRGAN       & StableSR        & RealBasicVSR        & RealViformer    & UAVideo         & MGLD            & \ours            \\ \hline\hline
    \multirow{7}{*}{REDS4}   & PSNR $\uparrow$        & \underline{25.92}   & 24.55     & 24.22             & 23.81           & 25.49               & \textbf{26.40}  & 24.32           & 25.57           & 25.11           \\
                             & SSIM $\uparrow$        & 0.6910              & 0.6875    & 0.6796            & 0.6514          & \underline{0.7069}  & 0.7063          & 0.6494          & 0.6781          & \textbf{0.7106} \\
                             & DISTS $\downarrow$     & 0.2571              & 0.1251    & 0.1335            & 0.1257          & 0.1120              & 0.1162          & 0.1506          & \textbf{0.0885} & \underline{0.1087}     \\
                             & tOF $\downarrow$       & 2.91                & 2.65    & 2.51          & 2.82            & \underline{2.09}    & 2.19            & 2.65            & 3.48            & \textbf{2.01}   \\
                             & $\text{tLP}^*$ $\downarrow$     & 1.97       & 2.26    & 1.53          & 4.12            & \underline{0.85}    & 1.26            & 1.65            & 3.01            & \textbf{0.71}   \\ 
                             & MUSIQ $\uparrow$       & 24.71               & 64.54    & 64.86        & \underline{67.31}      & 65.65        & 63.67           & 57.06           & 65.20           & \textbf{69.22}  \\
                             & $\text{DOVER}^*$ $\uparrow$     & 11.91    & 57.80    & 61.00        & 58.97          & 64.43        & 65.94           & 57.06           & \underline{68.83}      & \textbf{70.41}  \\ \hline
    \multirow{7}{*}{UDM10}   & PSNR $\uparrow$        & \underline{29.62}   & 28.46    & 27.93        & 26.69           & 28.94        & \textbf{30.41}  & 27.52           & 29.08           & 28.71           \\
                             & SSIM $\uparrow$        & 0.8536              & 0.8546    & 0.8446      & 0.7975          & 0.8542       & \textbf{0.8633} & 0.7975          & 0.8485          & \underline{0.8550}     \\
                             & DISTS $\downarrow$     & 0.1940              & 0.1230    & 0.1370      & 0.1565          & 0.1327       & 0.1274          & 0.1482          & \textbf{0.1168} & \underline{0.1220}     \\
                             & tOF $\downarrow$       & 1.38                & 1.28     & 1.33         & 1.45           & 1.11         & \underline{1.09}      & 1.34           & 1.53           & \textbf{1.04}  \\
                             & $\text{tLP}^*$ $\downarrow$     & 1.79       & 2.78    & 2.77          & 5.06            & 1.21         & \underline{1.00}       & 1.81            & 1.93            & \textbf{0.94}   \\ 
                             & MUSIQ $\uparrow$       & 28.06               & 61.41    & 61.44        & \underline{65.88}      & 63.43        & 60.86           & 64.19           & 62.79           & \textbf{67.16}  \\                & $\text{DOVER}^*$ $\uparrow$     & 15.73      & 74.29   & 76.02        & \underline{77.28} & 77.19        & 75.58           & 76.67           & 76.86      & \textbf{77.82}  \\ \hline

    \multirow{2}{*}{VideoLQ} 
                             & MUSIQ $\uparrow$       & 22.58               & 50.34    & 49.84        & 54.17           & \underline{55.96}   & 52.13           & 48.02           & 51.37           & \textbf{58.14}  \\
                             & $\text{DOVER}^*$ $\uparrow$     & 38.92      & 72.32    & 71.61        & 71.14           & 73.80        & 71.75           & 70.17           & \underline{73.98}      & \textbf{78.31}  \\ \Xhline{2\arrayrulewidth}
    \end{tabular}
    }
    \vspace{-2mm}
\end{table*}

\paragraph{DSSAG} To enhance high-frequency details, DSSAG offers a simpler approach without additional computational costs. The core idea of DSSAG is as follows.
As estimating noise from a noisy image is equivalent to estimating a noise-free image, we assume that the denoising U-Net of a diffusion model estimates a noise-free image in the following.
For estimating a noise-free image, the self-attention layers in a denoising U-Net find image regions with similar high-frequency details, by computing weights based on the similarities between queries and keys. Then, they aggregate information from different image regions based on their weights.
As noted by Wang~\etal~\shortcite{wang2018non}, this self-attention mechanism closely resembles bilateral filter~\cite{tomasi1998bilateral} and non-local means filter~\cite{buades2005nonlocal}, both of which are renowned structure-preserving filters.
Inspired by this, we introduce an additional parameter $\gamma$ to control the weighting function of the self-attention operation, similar to the weighting parameters in bilateral and non-local means filters.

\begin{figure}
    \centering
    \includegraphics[width=\linewidth]{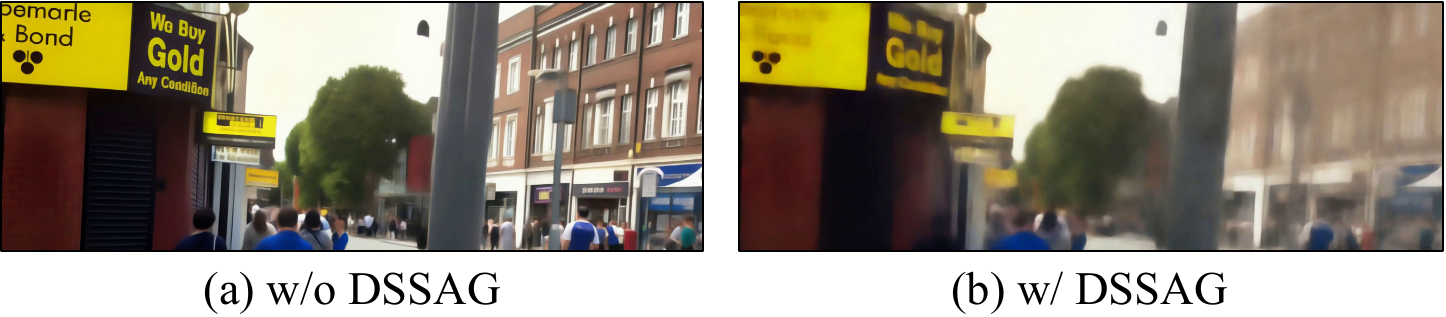}
    \vspace{-6mm}
    \caption{
    Denoised results of unconditional term at the intermediate timestep (t=16 out of 25) (a) without and (b) with DSSAG.
    The input video is from the VideoLQ dataset~\cite{chan2022investigating} (sample 013).
    }
    \vspace{-4mm}
    \label{fig:DSSAG}
\end{figure}

Specifically, we extend the self-attention operation as:
\begin{equation}
    SA(\bm{Q},\bm{K},\bm{V},\gamma) = \text{softmax}\left(\frac{\bm{Q}\bm{K}^\top}{\max(\gamma^2qk,1)\sqrt{d}}\right)\bm{V},
    \label{eq:SA_gamma}
\end{equation}
where $q$ and $k$ represent the largest absolute values among the elements of $\bm{Q}$ and $\bm{K}$, respectively.
We adopt $q$ and $k$ to adaptively control the weights to the scales of the keys and values. $\max(\cdot,1)$ is adopted to make \cref{eq:SA_gamma} reduce to the conventional self-attention operation, when $\gamma$ is small.
\cref{eq:SA_gamma} performs in a similar manner to the non-local means filter. Thanks to the similarity-based weighting function, it preserves salient image structures.
Moreover, $\gamma$ allows control over the blurring strength.
Assigning a large value to $\gamma$ results in larger weights for keys less similar to the queries, causing the information from different image regions to be more blended. Consequently, the denoising U-Net estimates a blurrier image with fewer high-frequency details as shown in \cref{fig:DSSAG}.

Leveraging the extended self-attention operation in \cref{eq:SA_gamma}, we define DSSAG and its combination with CFG as:
\begin{eqnarray}
    \epsilon_\text{DSSAG}(x_t)\!\!\!\!\!\!&=&\!\!\!\!\!\! \epsilon'_\theta(x_t)+(1+s)\left(\epsilon_\theta(x_t)-\epsilon'_\theta(x_t)\right)~~~\text{and}\\
    \epsilon_\text{CFG\&DSSAG}(x_t)\!\!\!\!\!\! &=&\!\!\!\!\!\! \epsilon'_\theta(x_t)+(1+s)\left(\epsilon_\theta(x_t,c)-\epsilon'_\theta(x_t)\right),
    \label{eq:CFG_DSSAG}
\end{eqnarray}
where $\epsilon'_\theta$ is a denoising U-Net whose self-attention operations are replaced with \cref{eq:SA_gamma}.
$\epsilon'_\theta$ does not require any training and shares the same parameters with $\epsilon_\theta$.
DSSAG offers a couple of distinct benefits compared to SAG~\cite{sag} and PAG~\cite{pag}.
It does not need additional high-frequency detection or blurring operations, as \cref{eq:SA_gamma} already incorporates these in its weighting and aggregation mechanism.
Furthermore, DSSAG provides smooth control over blur strength, unlike PAG, enabling seamless integration with CFG without any additional computational costs.

We apply DSSAG to the first two and last two spatial self-attention layers in the denoising U-Net, as done for SAP and TAP.
During the iterative sampling process of \ours{}, we set $\gamma$ adaptively to the noise level of the diffusion model, so that $\gamma$ is initially large and gradually decreases as the sampling proceeds.
Specifically, we set $\gamma_t$ at diffusion sampling step $t$ as:
\begin{equation}
    \gamma_t = \left(\frac{\ln{\sigma_t}-\ln{\sigma_T}}{\ln{\sigma_0}-\ln{\sigma_T}}\right)^{\rho}, 
\end{equation}
where $\sigma_t$ is the noise level at $t$, and $\rho$ is a parameter to control the detail suppression strength, which is set to $0.5$ in our experiments.
\section{Experiments}
To build \ours{}, we fine-tune Image-to-Video Stable Video Diffusion (I2V-SVD) \cite{svd}, which adopts the LDM framework~\cite{ldm} with the EDM~\cite{edm} diffusion mechanism.
We use the REDS dataset~\cite{reds} to train our model. Following previous work \cite{chan2022investigating,yang2023mgldvsr}, we merge 240 training videos and 30 test videos, reorganizing them into 266 training videos and 4 test videos, and refer to the latter as REDS4.
We refer the reader to the supplementary material for more implementaion details.

We use the REDS4 \cite{chan2022investigating} and UDM10 \cite{udm10} datasets for VSR evaluation.
We construct HR-LR video pairs from the datasets using the real-world degradation pipeline of Chan \etal~\shortcite{chan2022investigating}, which applies random blur, resizing, noise, JPEG compression, and video compression.
Additionally, we use the VideoLQ dataset~\cite{chan2022investigating} as a real-world LR dataset.

\subsection{Comparison with Previous SR and VSR Approaches}
To evaluate the performance of {\ours}, we compare it with various previous methods across different categories.
Specifically, our evaluation includes non-generative image super-resolution (ISR) methods: bicubic interpolation and SwinIR~\cite{liang2021swinir}; two generative ISR methods: Real-ESRGAN~\cite{real-esrgan} and StableSR~\cite{stablesr}; two non-generative VSR methods: RealBasicVSR~\cite{chan2022investigating} and RealViformer~\cite{realviformer}; and two generative VSR methods: Upscale-A-Video (UAVideo)~\cite{zhou2024upscaleavideo} and MGLD~\cite{yang2023mgldvsr}.
We follow the default settings from the official repositories of prior works in our experiments.

For quantitative evaluation on synthetic datasets that provide ground-truth HR videos, we use PSNR, SSIM, DISTS~\cite{dists}, tOF, and tLP~\cite{tOFtLP}.
PSNR, SSIM, and DISTS measure the quality of each frame in super-resolution results, while tOF and tLP measure the temporal consistency by comparing optical flows estimated from super-resolution results with ground-truth optical flows.
We also conduct a quantitative evaluation on a real-world dataset, VideoLQ~\cite{chan2022investigating}, using no-reference quality metrics: MUSIQ~\cite{musiq} and DOVER~\cite{dover}.
MUSIQ measures perceptual image quality, while DOVER is a video quality metric that quantifies aesthetic quality and technical artifacts such as compression or blur.

\begin{figure}[t]
    \centering
    \includegraphics[width=0.95\linewidth]{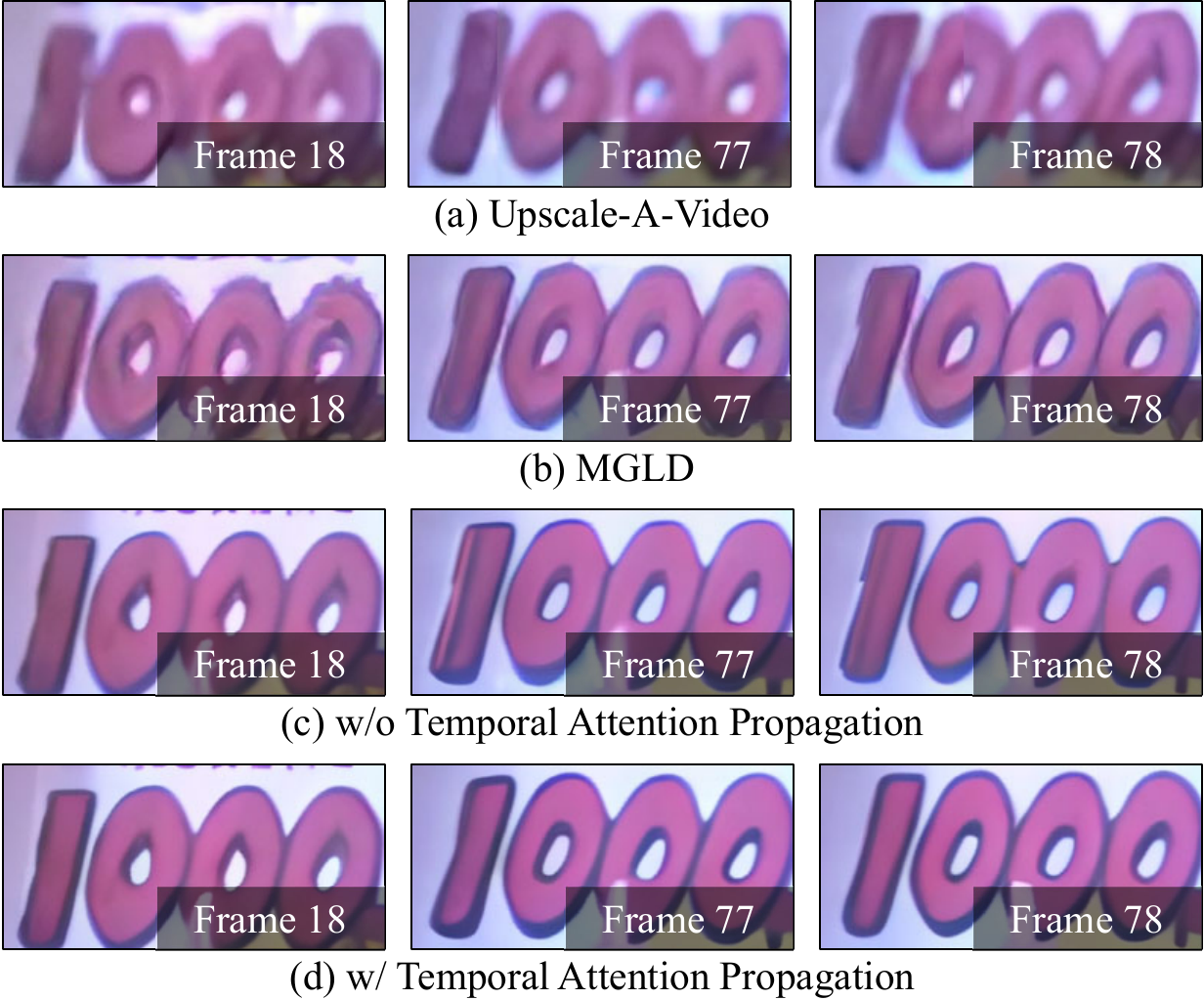}
    \vspace{-2mm}
    \caption{(a)\&(b) are VSR results of image diffusion prior-based methods. (c)\&(d) show the effects on the proposed {\tfiabb}.
    The input video is from the VideoLQ dataset~\cite{chan2022investigating} (sample 008).
    }
    \label{fig:temporal_ablation}
    \vspace{-2mm}
\end{figure}

\cref{table:1} presents a quantitative comparison with previous methods. It is important to note that quantitatively measuring VSR quality is challenging, especially for generative-prior-based approaches, due to the synthesized details that may not align with the ground-truth details. Nevertheless, on both synthetic and real-world datasets, \ours{} consistently demonstrates superior or comparable results across various metrics. For PSNR and SSIM, non-generative methods generally achieve higher scores as they do not synthesize high-frequency details that may not align with ground-truth videos. 
However, this often results in blurry outputs as will be shown in   \cref{fig:spatial_qualitative,fig:temporal_qualitative,fig:fo1}.
Compared to previous generative VSR methods, \ours{} achieves competitive PSNR scores and higher SSIM scores.
When considering DISTS, a perceptual metric, \ours{} ranks second among all methods. Regarding temporal consistency, measured by tOF and tLP, \ours{} outperforms all other methods, including non-generative ones, highlighting the effectiveness of video generative prior and our {\tfiabb} scheme. Finally, evaluating video quality independent of ground-truth data using MUSIQ and DOVER, \ours{} outperforms other methods in most cases on both synthetic and real-world datasets.

\cref{fig:spatial_qualitative} presents a qualitative comparison focusing on spatial consistency. 
Among the compared methods, RealViformer~\cite{realviformer} processes the entire spatial area of video frames without splitting them into multiple spatial tiles. However, it lacks mechanisms, such as spatial attention, to ensure consistency between distant regions, resulting in spatially inconsistent artifacts. The other methods, including ours, adopt tile-based approaches. 
For all tile-based approaches, the regions marked by red squares in the LR frames belong to different tiles.
As shown in the results, our method produces high-quality, spatially-consistent results despite our tile-based approach, surpassing the other approaches.

\cref{fig:temporal_qualitative} presents a qualitative comparison focusing on temporal consistency. In the figure, all previous approaches exhibit challenges in producing temporally consistent HR details for both distant and neighboring frames. StableSR~\cite{stablesr}, an image super-resolution (ISR) approach, handles each video frame separately, which inherently limits its temporal consistency.
RealViformer~\cite{realviformer} propagates information bidirectionally between neighboring frames, but this method still struggles with ensuring consistency across longer temporal sequences.
Moreover, due to the lack of generative prior, its results show less realistic details.
Meanwhile, MGLD~\cite{yang2023mgldvsr} and UAVideo~\cite{zhou2024upscaleavideo}, both VSR approaches, rely on image diffusion priors and process temporal tiles, but struggle to maintain temporal consistency.
Finally, our results show superior temporal consistency and higher-quality HR details than the results of previous methods.

\cref{fig:fo1} presents a qualitative comparison highlighting the restoration quality of a single frame, demonstrating the superiority of {\ours} over previous methods. For example, {\ours} excels at restoring building windows, adding plausible details to degraded regions, as seen in VideoLQ 020.
Notably, it surpasses other VSR methods in restoring characters, a particularly challenging task for generative models, as seen in VideoLQ 049.
These results indicate that {\ours} not only ensures spatio-temporal consistency but also achieves outstanding restoration quality.

The superior performance of {\ours}, demonstrated in \cref{fig:spatial_qualitative,fig:temporal_qualitative,fig:fo1}, stems from the video diffusion prior and our proposed components.
Specifically, the video diffusion prior leverages spatially and temporally distant information within a spatio-temporal tile, ensuring not only spatio-temporal consistency but also enhanced HR details.
Furthermore, our SAP and TAP schemes allow exploiting information from distant tiles, further improving spatio-temporal consistency and HR details.
Finally, our DSSAG scheme additionally improves the quality of HR details.

\begin{figure}[t]
    \centering
    \includegraphics[width=0.95\linewidth]{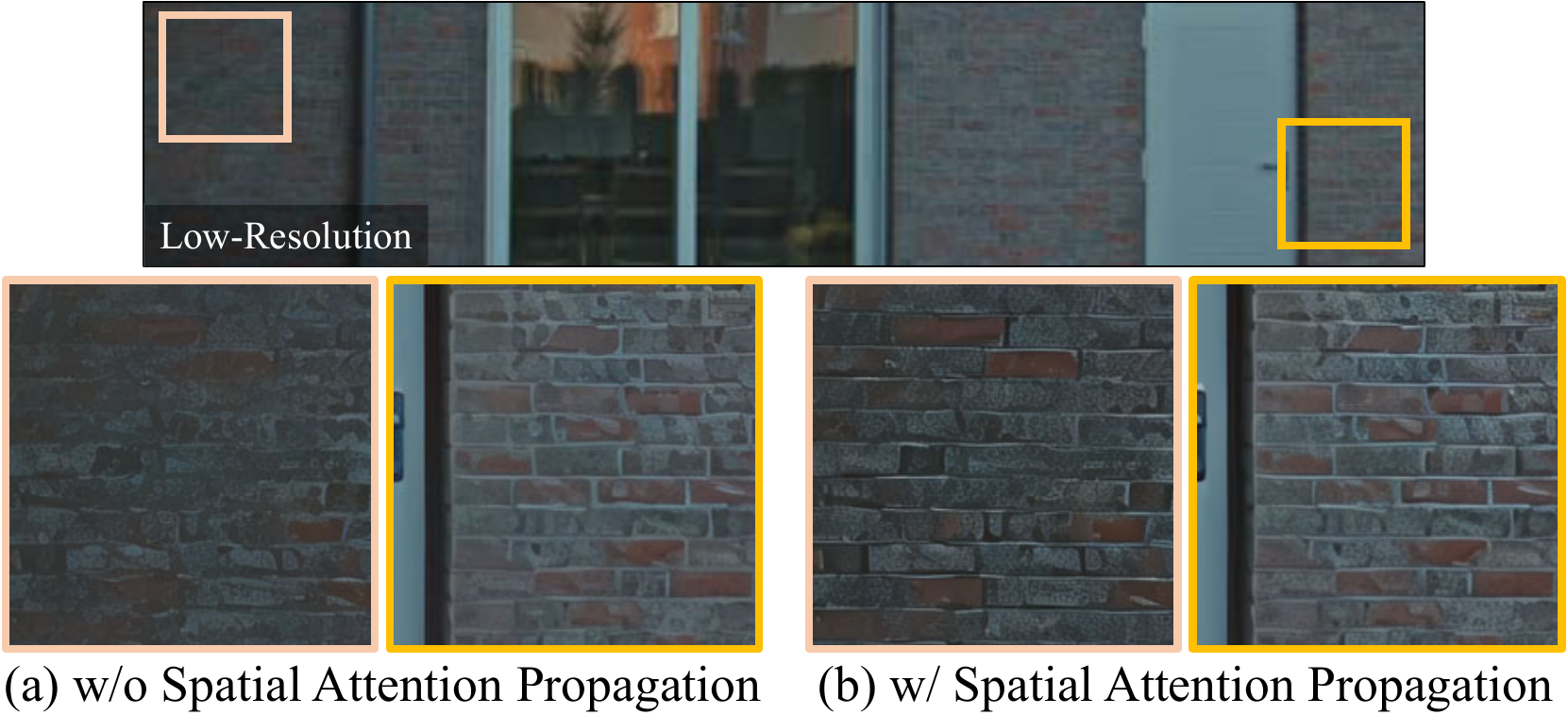}
    \vspace{-2mm}
    \caption{Visualization of the effects on the proposed {\ssiabb}.
    The input video is from Pexels (\copyright Tima Miroshnichenko).
    }
    \label{fig:spatial_ablation}
    \vspace{-5mm}
\end{figure}

\subsection{Component Evaluation}
\label{ssec:ablation_study}

\paragraph{Effect of Video Diffusion Prior}
\cref{fig:temporal_ablation}(a-b) and (c) show the VSR results of previous methods utilizing image diffusion priors and our method employing video diffusion priors, respectively. While image diffusion prior-based methods incorporate additional techniques to enhance temporal consistency, they produce blurry and inconsistent outcomes. In contrast, our approach, which directly applies video diffusion priors without additional temporal consistency techniques, shows significantly clearer and more temporally consistent results.

\paragraph{Ablation Study on {\tfiabb} \& {\ssiabb}} 
\cref{fig:temporal_ablation}(c) and (d) present the VSR results without and with {\tfiabb}, respectively. By applying {\tfiabb}, temporal consistency between distant frames is effectively preserved. Specifically, in frame 77, a vertical line in the number ``1'' appears in (c) but remains absent in (d), maintaining the original consistency. 
\cref{fig:spatial_ablation} illustrates the VSR results of two distant brick regions in a frame, comparing outcomes without and with {\ssiabb}, respectively. Without {\ssiabb}, the brick patterns appear mismatched between the regions. In contrast, after applying {\ssiabb}, the distant brick patterns are restored with coherent details.
Additionally, the first three rows in \cref{tab:ablation} present ablation studies on {\ssiabb} and {\tfiabb}. Both schemes show improvements in the DOVER. In contrast, since MUSIQ measures image quality, no improvement is observed for {\tfiabb}. For a detailed visualization of the enhancements brought by SAP and TAP, please refer to the supplemental video.

\begin{figure}[t]
    \centering
    \includegraphics[width=0.95\linewidth]{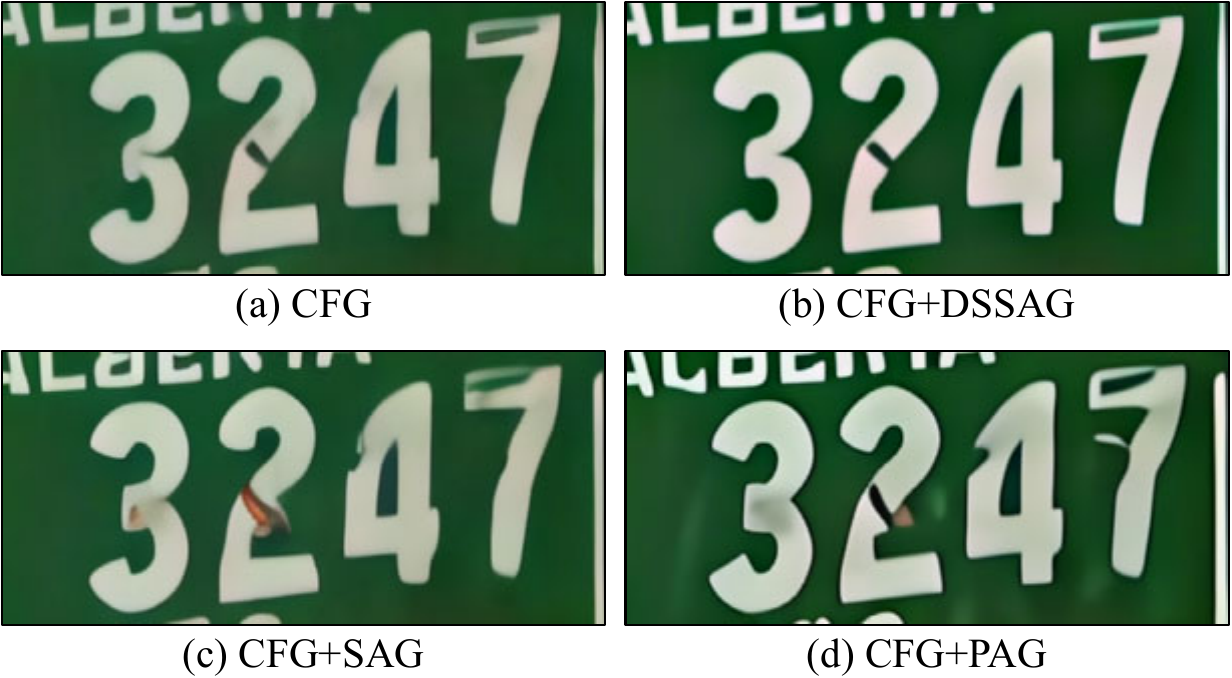}
    \vspace{-2mm} 
    \caption{Qualitative comparison with previous guidance approaches.
    The input video is from the VideoLQ dataset~\cite{chan2022investigating} (sample 041).
    }
    \vspace{-5mm}
    \label{fig:ablation_dssag}
\end{figure}

\begin{figure*}[t]
    \centering
    \vspace{-4mm}
    \includegraphics[width=\linewidth]{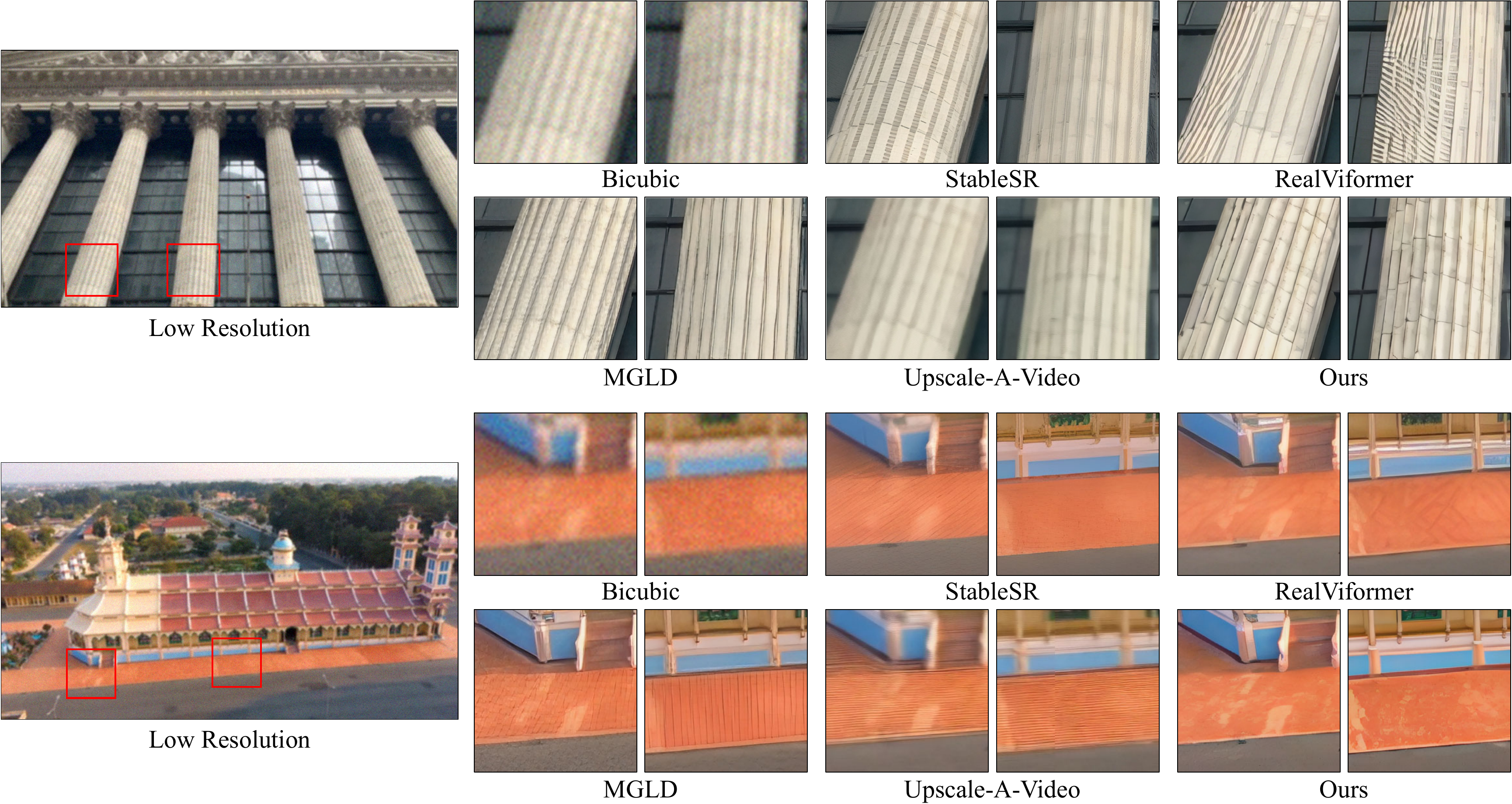}
    \caption{Qualitative comparison of VSR (x4) with other state-of-the-art methods. Our proposed {\ssi} ({\ssiabb}) restores two spatial regions with consistent details and high quality.
    The input videos are from Pexels (\copyright CityXcape, \copyright lam loi).
    }
    \Description{spatial qualitative}
    \label{fig:spatial_qualitative}
\end{figure*}

\begin{figure*}[t]
    \centering
    \includegraphics[width=\linewidth]{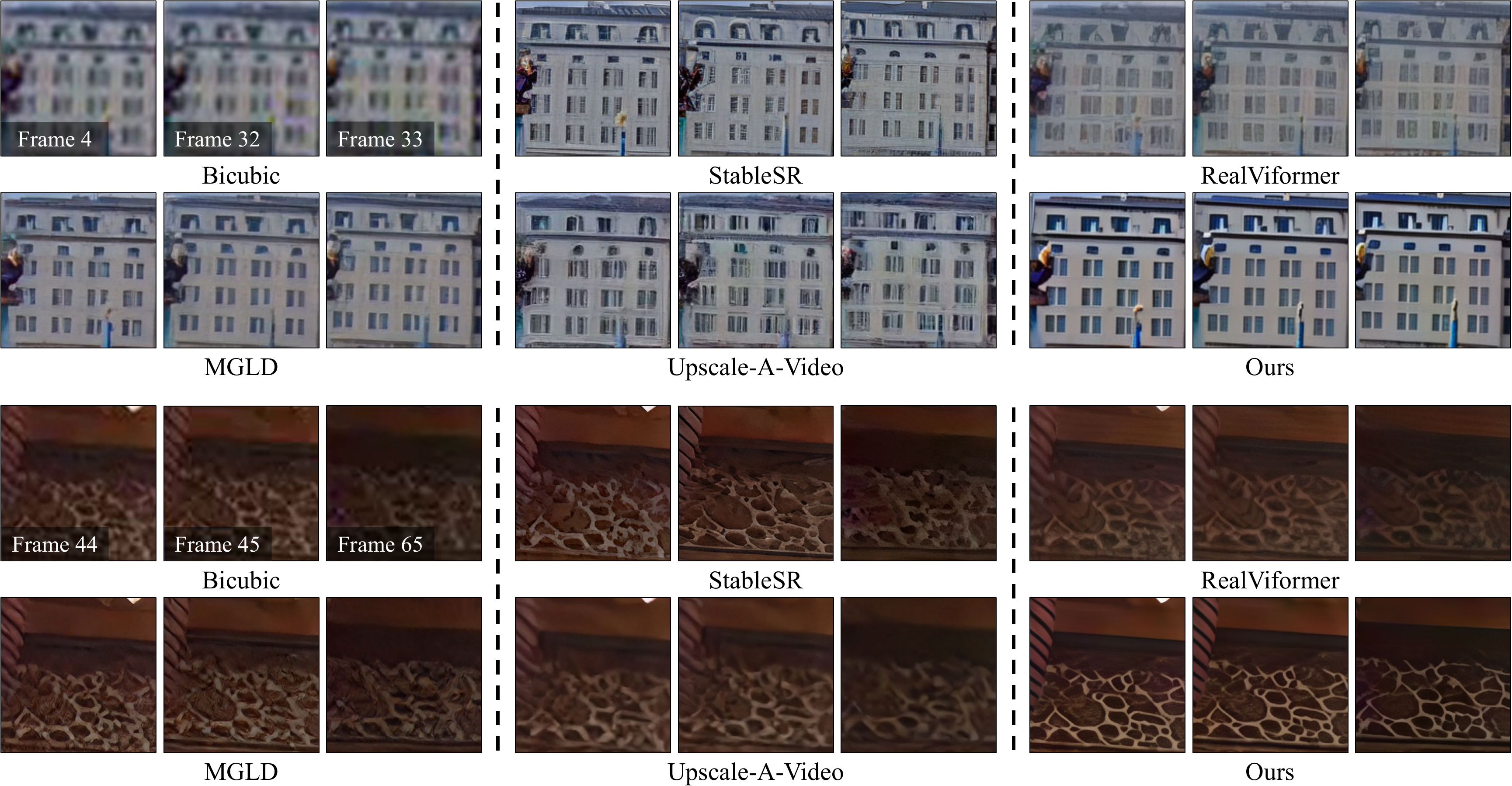}
    \vspace{-4mm}
    \caption{Qualitative comparison of VSR (x4) with other state-of-the-art methods. Our proposed {\tfi} ({\tfiabb}) restores consecutive and distant frames with consistent details and high quality.\
    The input videos are from the VideoLQ dataset~\cite{chan2022investigating} (samples 007 and 036).
    }
    \Description{temporal qualitative}
    \label{fig:temporal_qualitative}
\end{figure*}

\begin{figure*}[t]
    \centering
    \includegraphics[width=\linewidth]{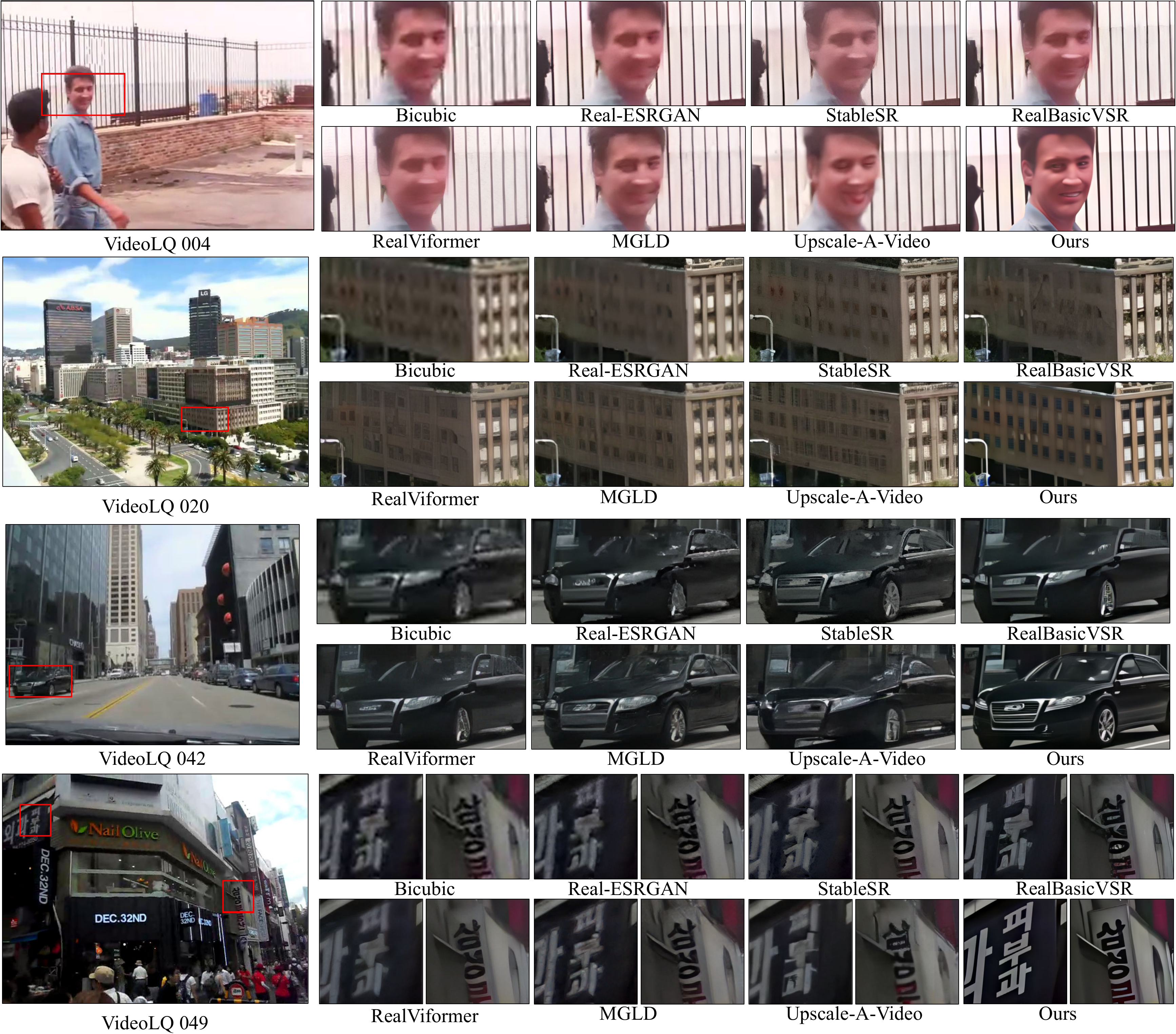}
    \vspace{-6mm}
    \caption{Qualitative comparison of VSR (x4) on the real-world video dataset (VideoLQ)~\cite{chan2022investigating} with other state-of-the-art methods.
    The input videos are from the VideoLQ dataset~\cite{chan2022investigating}.
    }
    \vspace{-2mm}
    \Description{overall qualitative}
    \label{fig:fo1}
\end{figure*}

\paragraph{Comparisons with Guidance Methods}
\cref{fig:ablation_dssag} presents qualitative comparisons of {\drgabb} with previous guidance approaches, where our method shows the clearest image quality and most accurate character shape.
The last three rows in \cref{tab:ablation} provide quantitative comparisons, showing that {\drgabb} achieves the best performance in DOVER and the second-best in MUSIQ, while being $1.5\times$ faster than previous methods. Moreover, since these guidance methods can be generally applicable beyond VSR, we provide additional comparisons for image generation in the supplemental document.

\begin{figure*}[t]
    \centering
    \includegraphics[width=\linewidth]{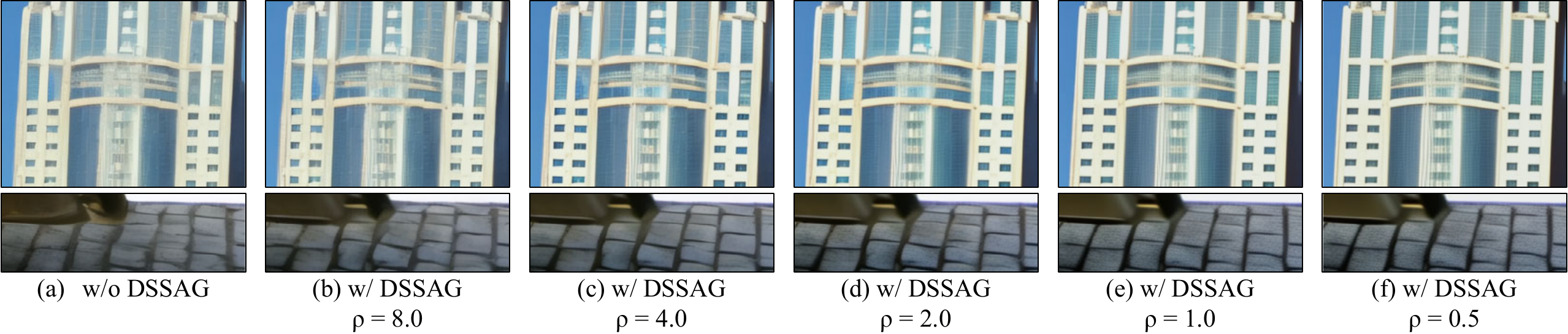}
    \vspace{-6mm}
    \caption{Effects of $\rho$ in {\drgabb}. (a) and (b)-(f) are the results of {\ours} without and with {\drgabb}. The ability to refine and enhance high-frequency increases as the $\rho$ value decreases.
    The input videos are from the VideoLQ dataset~\cite{chan2022investigating} (samples 024 and 037).
    }
    \Description{DSSAG rho}
    \label{fig:dssag_rho}
\end{figure*}

\paragraph{Effect of $\rho$ in {\drgabb}}
{\drgabb} adjusts the detail suppression strength using $\rho$.
\cref{fig:dssag_rho} shows the effect of $\rho$.
A smaller $\rho$ increases the blurring effect in \cref{eq:SA_gamma}, resulting in more enhanced edges. In contrast, a larger $\rho$ makes \cref{eq:SA_gamma} closer to the conventional self-attention operation, producing less enhanced but more accurate details. This flexibility allows our framework to adapt to various scenarios: a small $\rho$ is suitable for severely degraded LR videos or animation videos, while a large $\rho$ works better for mildly degraded LR videos.

\begin{table}
  \caption{Ablation study of {\ssiabb} \& {\tfiabb} and comparison with previous guidance methods using the REDS4 dataset. $\text{\normalfont DOVER}^*$ denotes {\normalfont DOVER ($\times$100)}. The best and second-best scores are marked in \textbf{bold} and \underline{underline}. Our proposed {\drgabb} is $1.5\times$ faster than SAG and PAG.}
  \scalebox{0.83}{
  \label{tab:ablation}
  \begin{tabular}{ccccc|cc}
    \Xhline{2\arrayrulewidth}
    {\ssiabb} & {\tfiabb} & {\drgabb} & {SAG~\shortcite{sag}} & {PAG~\shortcite{pag}} & {MUSIQ $\uparrow$} & {$\text{DOVER}^*$ $\uparrow$} \\
    \hline\hline
    - & - & - & - & - & 67.51 & 66.07 \\
    \ding{51} & - & - & - & - & 67.52 & 66.10 \\
    - & \ding{51} & - & - & - & 67.51 & 66.19 \\
    \midrule
    \ding{51} & \ding{51} & \ding{51} & - & - & \underline{69.22} & \textbf{70.41} \\
    \ding{51} & \ding{51} & - & \ding{51} & - & 68.30 & \underline{69.43}\\
    \ding{51} & \ding{51} & - & - & \ding{51} & \textbf{71.24} & 68.55 \\
    \Xhline{2\arrayrulewidth}
  \end{tabular}
  }
\end{table}


\section{Conclusion}
In this paper, we introduced \ours{}, a novel approach leveraging a video diffusion prior to produce spatially and temporally consistent VSR results with realistic textures. Our proposed SAP and TAP schemes effectively address the challenges of maintaining consistency across spatio-temporal tiles. Furthermore, our DSSAG provides high-frequency detail enhancement without incurring additional computational costs.

\paragraph{Limitations \& Future Work}
Since \ours{} is based on a diffusion model, it may generate false details and face difficulties in achieving real-time VSR. Moreover, \ours{} may struggle to maintain temporal consistency across distant temporal tiles in long video clips, as TAP primarily relies on nearby tiles. Addressing these issues could be a promising future direction.

\begin{acks}
This work was supported by Samsung Electronics Co., Ltd., the Institute of Information \& Communications Technology Planning \& Evaluation (IITP) grants (RS-2019-II191906, Artificial Intelligence Graduate School Program (POSTECH), No.2021-0-02068, Artificial Intelligence Innovation Hub), and the National Research Foundation of Korea (NRF) grant (No.2018R1A5A1060031), funded by the Korea government (MSIT).
\end{acks}

\bibliographystyle{ACM-Reference-Format}
\bibliography{sections/reference}

\clearpage
\twocolumn

\appendix
\section*{Supplementary Material}

We uploaded a \href{https://www.youtube.com/watch?v=EaHrA9bcMSs}{\textcolor{blue}{video}} as Supplementary Material, containing VSR results and comparisons that include the contents mentioned in the main paper.

\section{Implementation Details}
We fine-tune Image-to-Video Stable Video Diffusion (I2V-SVD) \cite{svd} with an $8\times$ downsampling VAE \cite{ldm}, trained on 14-frame sequences at a resolution of $576\times1024$.
We train {\ours} using the AdamW optimizer \cite{loshchilov2019decoupled} with a learning rate of $1 \times 10^{-4}$ and a batch size of 256 on four NVIDIA A100-80GB GPUs. Due to the memory limit, we use gradient accumulation to match the size of the batch. The training data are cropped into spatio-temporal tiles with width, height, and sequence length set to 512, 512, and 14, respectively. The denoising U-Net receives motion magnitude and a frame rate, fixed at 127 and 8, respectively, as conditioning inputs. 
Additionally, the denoising U-Net takes a CLIP \cite{clip} embedding vector of the first frame of the low-resolution input video as a text prompt.
We initialize VAE and U-Net network parameters to pre-trained I2V-SVD \cite{svd} parameters. During training, we freeze the VAE and fine-tune all U-Net parameters for the first 5K iterations. Afterward, we fix the temporal attention and temporal residual block layers in the U-Net and fine-tune the U-Net for additional 5K iterations.

In our framework, {\ssi} ({\ssiabb}), {\tfi} ({\tfiabb}) and {\drg} ({\drgabb}) are adopted to every spatial self-attention layer in the first two down-blocks and the last two up-blocks in the denoising U-Net. Merged keys and values in {\ssiabb} and {\tfiabb} are computed by concatenating each key and value along a flattened spatial dimension in the self-attention mechanism.
In our experimental settings, {\ours} is capable of performing video super-resolution ($\times4$) on input videos of up to 768$\times$768 resolution with 21 frames in a single NVIDIA A100-80GB GPU.

\section{Alternating and Simultaneous Strategies for {\ssiabb} and {\tfiabb}}
While \ssiabb{} and \tfiabb{} may be applied simultaneously,  our approach adopts an alternating application strategy to improve computational efficiency. 
\cref{tab:altvssim} provides a performance comparison of the two approaches on the VideoLQ \cite{chan2022investigating} dataset.
Both approaches show similar performance in MUSIQ \cite{musiq} and $\text{\normalfont DOVER}^*$ \cite{dover}, but the simultaneous approach requires twice the inference time of the alternating approach.  Therefore, we adopt the alternate approach in our proposed {\ours} to ensure computational efficiency.

\begin{figure}[t]
    \centering
    \includegraphics[width=\linewidth]{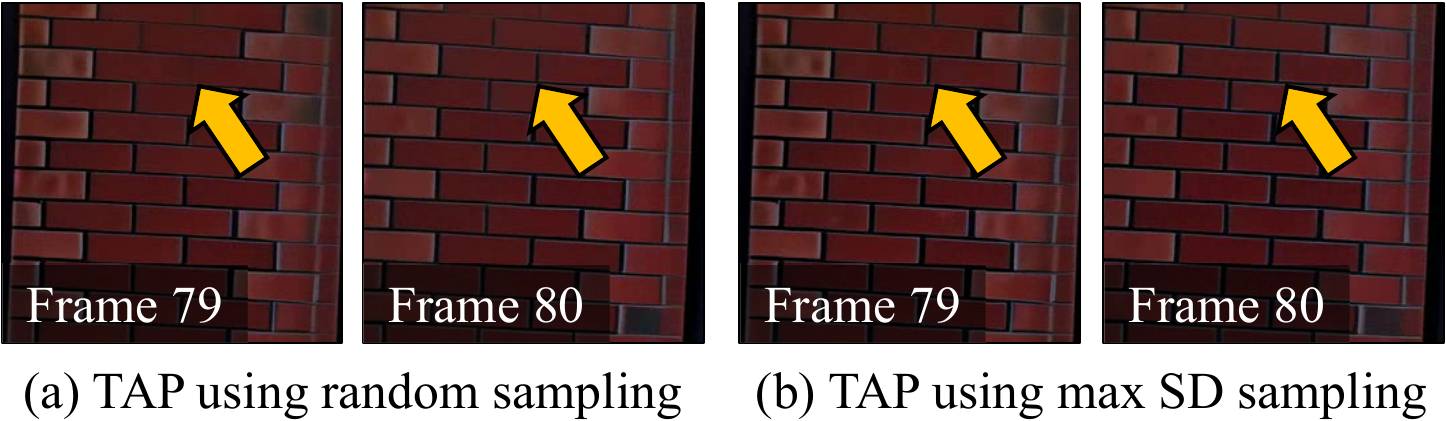}
    \caption{Comparison of two different {\tfiabb} sampling strategy.
    The input video is from the VideoLQ dataset~\cite{chan2022investigating} (sample 013).
    }
    \label{fig:tfi_sampling}
\end{figure}

\begin{table}[t]
  \caption{
  Comparison of alternating and simultaneous application of {\ssiabb} and {\tfiabb} on the VideoLQ \cite{chan2022investigating} dataset. The best and second-best scores are marked in \textbf{bold} and \underline{underline}.
  }
  \label{tab:altvssim}
  \begin{tabular}{c|cc}
    \Xhline{2\arrayrulewidth}
        Approach & {Alternating} & {Simultaneous} \\
        \hline\hline
        MUSIQ $\uparrow$ & \textbf{58.15} & \underline{58.14} \\
        $\text{DOVER}^*$ $\uparrow$ & \underline{78.30} & \textbf{78.31} \\
    \Xhline{2\arrayrulewidth}
  \end{tabular}
\end{table}

\section{Sampling Strategy for {\tfiabb}}
In this section, we show the effectiveness of our sampling strategy for {\tfiabb}. Within the {\tfiabb} scheme, we reduce the computational cost by selecting 4 frames from the previous spatial tile, which consists of 14 frames. A straightforward approach is to select frames randomly. However, we opt for a more informed strategy: selecting frames whose keys exhibit large standard deviations. This choice is based on the observation that frames with more varied and sharper details tend to produce distinct keys, resulting in larger standard deviations. \cref{fig:tfi_sampling} illustrates this trend by comparing random sampling with maximum standard deviation (max SD) sampling. As shown, max SD sampling produces consistent and detailed brick patterns, in contrast to random sampling. This demonstrates that selecting detail-rich frames based on their standard deviations enhances the quality and consistency of temporal tiles.

\section{Tile Propagation Range of {\tfiabb}}
In {\tfiabb}, temporal information is propagated bidirectionally from both the previous and next tiles. While increasing the propagation range can further enhance temporal consistency, it also introduces higher computational overhead and raises the risk of propagating inconsistent information.
\cref{tab:taprange} compares the impact of different propagation ranges in terms of warping error (WE). The results indicate that increasing the propagation range does not lead to significant performance improvements, as the directly adjacent tiles already provide sufficient information for {\tfiabb}.
Furthermore, although a propagation range of 2 achieves the best performance, further extending the range degrades performance due to the propagation of inconsistent information between distant tiles. In our experiments, we adopt a propagation range of 1, as the performance improvement from additional tiles is minimal compared to the substantial increase in memory usage, requiring 3.2GB of GPU memory per tile.

\begin{table}[t]
  \caption{
  Warping error (WE) as a function of the tile propagation range of TAP on the VideoLQ \cite{chan2022investigating} datasets (no.40-49). The best and second-best scores are marked in \textbf{bold} and \underline{underline}, respectively.
  }
  \label{tab:taprange}
  \begin{tabular}{c|cccc}
    \Xhline{2\arrayrulewidth}
        \# Tiles & 1 & 2 & 3 & 4 \\
        \hline\hline
        WE $\downarrow$ & {0.1749} & \textbf{0.1744} & \underline{0.1745} & {0.1746}\\
    \Xhline{2\arrayrulewidth}
  \end{tabular}
\end{table}

\section{Subsampling Rates of {\ssiabb} and {\tfiabb}}
{\ssiabb} and {\tfiabb} leverage subsampled information from the current frame and previous temporal tile to improve spatial and temporal consistency.
A low subsampling rate, which preserves more information, can lead to improved performance, but significant computational costs are required.
\cref{tab:srsap} shows correlation between subsampling rates and performance changes in {\ssiabb}.
We compare the average similarity over 28 frames between two distant spatial tiles, as indicated in Fig.~5 of our main paper.
As the subsampling rate decreases, LPIPS~\cite{lpips}, ST-LPIPS~\cite{stlpips}, and CLIP similarity~\cite{clip} scores improve, indicating the performance increase. Consequently, utilizing the full information from the current frame (100\% in \cref{tab:srsap}) yields the best results.

\cref{tab:srtap} shows warping error (WE) under varying subsampling rates of {\tfiabb} on the VideoLQ \cite{chan2022investigating} dataset (no.40-49).
Similar to the trend in {\ssiabb}, sampling more key-value pairs, which correspond to more extensive use of previous tile information, enhance temporal consistency.
In our experimental settings, we select the subsampling rate of 2 for {\ssiabb}, and 4 key-value pairs for {\tfiabb}, which correspond to $50\%$ and $28\%$ sampling, respectively, considering computational overhead and performance improvement.

\section{Ablation Study of {\drgabb}}
\begin{figure}[t]
    \centering
    \includegraphics[width=\linewidth]{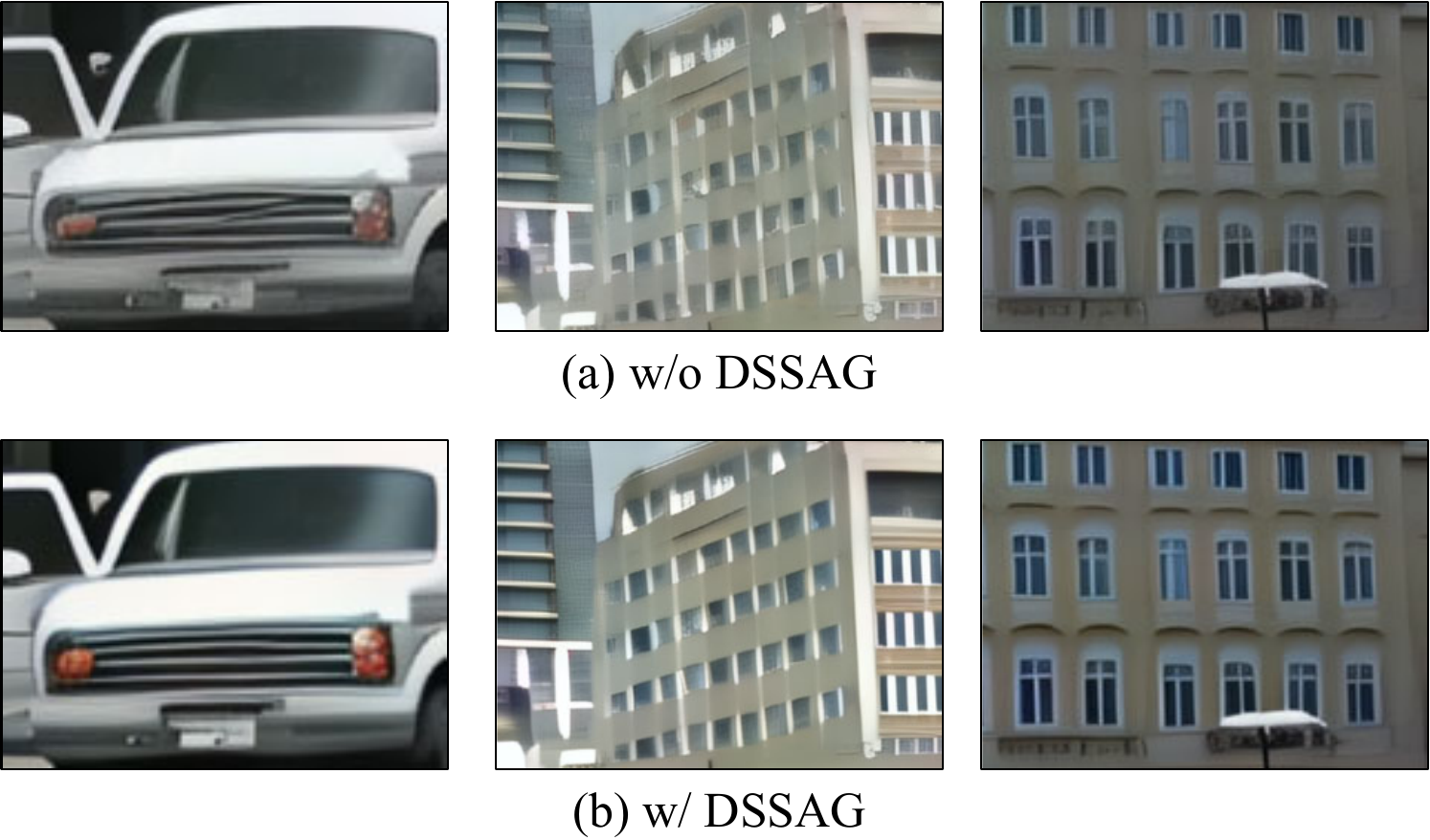}
    \caption{Effects of {\drgabb}. The first row and second row 
     are the results of {\ours} without and with {\drgabb}. The proposed method enhances the generation quality of video frames.
     The input videos are from the VideoLQ dataset~\cite{chan2022investigating} (sample 024, 039, and, 042).
     }
    \Description{ablation figures}
    \label{fig:ablation_drg}
\end{figure}

\cref{fig:ablation_drg} demonstrates the effectiveness of {\drgabb} by comparing VSR results with and without {\drgabb}. Without {\drgabb} in \cref{fig:ablation_drg}, the car grille and the bases of the windows exhibit significant distortions. However, after applying {\drgabb}, these elements are restored with clear and precise details. These results highlight the effectiveness of {\drgabb} in enhancing visual quality and structural fidelity in VSR.

\begin{table}[t]
  \caption{
    Performance comparison based on {\ssiabb} subsampling rates. The best and second-best scores are marked in \textbf{bold} and \underline{underline}, respectively.
  }
  \label{tab:srsap}
  \begin{tabular}{c|cccc}
    \Xhline{2\arrayrulewidth}
        Subsampling rate & 8 (12.5\%) & 4 (25\%) & 2 (50\%) & 1 (100\%) \\
        \hline\hline
        LPIPS $\downarrow$ & 0.4693 & 0.4551 & \underline{0.4416} & \textbf{0.4303} \\
        ST-LPIPS $\downarrow$ & 0.3065 & 0.2846 & \underline{0.2641} & \textbf{0.2598} \\
        CLIP Similarity $\uparrow$ & 0.9500 & 0.9574 & \underline{0.9618} & \textbf{0.9637} \\
    \Xhline{2\arrayrulewidth}
  \end{tabular}
\end{table}

\begin{table}[t]
  \caption{
    Performance comparison based on {\tfiabb} subsampling rates.
    The top row represents the number of subsampled key-value pairs out of 14 key-value pairs from the previous temporal tile.
    The best and second-best scores are marked in \textbf{bold} and \underline{underline}, respectively.
  }
  \label{tab:srtap}
  \begin{tabular}{c|cccc}
    \Xhline{2\arrayrulewidth}
        \# key-value pairs & 1 & 2 & 4 & 8 \\
        \hline\hline
        WE $\downarrow$ & 0.1767 & 0.1758 & \underline{0.1749} & \textbf{0.1742} \\
    \Xhline{2\arrayrulewidth}
  \end{tabular}
\end{table}

\section{Image Generation Comparison of {\drgabb}}
\begin{figure*}[t]
    \centering
    \scalebox{0.8}{
    \includegraphics[width=\textwidth]{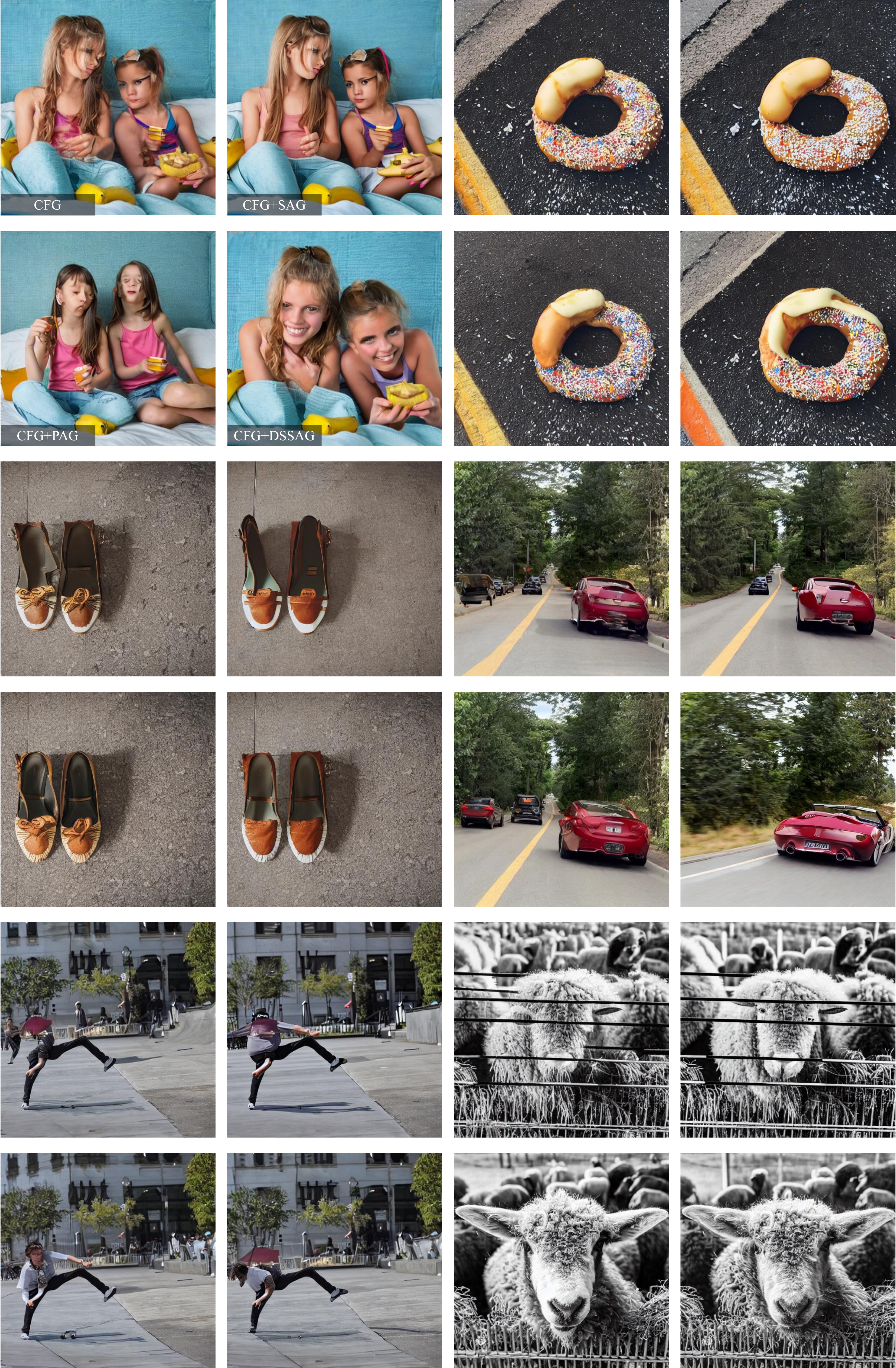}
    }
    \caption{Visualization of previous diffusion guidance approaches and our {\drgabb}. {\drgabb} combines well with CFG, while 1.5$\times$ faster than SAG~\shortcite{sag} and PAG~\shortcite{pag}.
    Images are generated using a diffusion model~\cite{sd}.
    }
    \label{fig:qual_dssag_sd}
\end{figure*}

The guidance methods, such as SAG \cite{sag}, PAG \cite{pag}, and our {\drgabb}, are generally applicable to the diffusion sampling process. To demonstrate the general effectiveness of {\drgabb}, we compare the proposed method with SAG and PAG in the image generation task. Specifically, we generate 10K images using Stable Diffusion \cite{sd} and evaluate them using the Inception Score (IS) \cite{is} and Fr\'echet Inception Distance (FID) \cite{fid} metrics. The MS-COCO 2014 validation dataset \cite{coco} is used for these measurements. \cref{tab:image_synthesis} presents the comparison. The proposed method achieves the best FID score and a comparable IS score, highlighting the general effectiveness of {\drgabb}. Notably, this performance is achieved with a $1.5\times$ faster inference speed. \cref{fig:qual_dssag_sd} shows visualization of previous diffusion guidance approaches and {\drgabb}. {\drgabb} effectively improves image synthesis quality and integrates well with CFG.

\section{Inference Time Comparison}
\cref{tab:inferencetime} presents a computation time comparison of diffusion-based video super-resolution methods for $4\times$ super-resolution of $640\times360$ videos.
Although {\ours} significantly outperforms the previous approaches, it incurs slightly higher computational overhead, mainly due to the need for spatio-temporal tile splitting and merging at each diffusion timestep and attention-based approaches for maintaining spatial and temporal consistency.

\section{Image-to-Video Stable Video Diffusion}
I2V-SVD employs LDM \cite{ldm} structure with EDM \cite{edm} diffusion mechanism, which downsamples (x8) input with VAE encoder $\mathcal{E}$ and proceeds continuous time diffusion denoising process in latent space. After the denoising process ended, it upsamples (x8) latent vectors with VAE decoder $\mathcal{D}$ to the original RGB space. For training EDM with the LDM structure, diffusion loss can be defined as:
\begin{equation}
  L(\theta) = \mathbb{E}_{\sigma,\bm{y},\bm{n}} \left[ \lambda(\sigma) || D_{\theta}(\bm{y+n}, \bm{c}, \sigma) - \bm{y} ||^{2}_{2} \right],
  \label{eq:1}
\end{equation}
where $D_{\theta}$ is a denoiser network, $\bm{c}$ is additional conditions such as text prompt, motion magnitude, and frame rate, $\sigma$ is sampled noise level from predefined $p_{train}(\sigma)$, $\bm{y} \sim p_{\mathcal{E}(data)}$, $\bm{n} \sim \mathcal{N}(\bm{0},\sigma^2\bm{I})$ and $\lambda(\sigma)$ is adaptive loss weight. Here, CLIP \cite{clip} embedding vectors of the input image is given as text prompt. For stable and consistent input and output signal magnitudes, $D_{\theta}$ is derived from network $F_{\theta}$ in the following form:
 \begin{equation}
     D_{\theta}(\bm{x}, \bm{c}, \sigma)=c_{\text{skip}}(\sigma) \ \bm{x}+c_{\text{out}}(\sigma) \ F_{\theta}(c_{\text{in}}(\sigma) \ \bm{x}, \bm{c}, c_{\text{noise}}(\sigma)),
     \label{eq:2}
 \end{equation}
where $c_\text{skip}(\sigma)$, $c_\text{out}(\sigma)$, $c_\text{in}(\sigma)$ and $c_\text{noise}(\sigma)$ are parameterized function over the noise level $\sigma$ and $\bm{x}=\bm{y}+\bm{n}$. In the sampling stage, noise levels $\sigma_{0\sim T}$ are selected in the range $[0.002, 700]$ in descending order, and the deterministic diffusion backward process is calculated by solving ordinary differential equations from $\bm{x}_{0} \sim \mathcal{N}(\bm{0}, 700^2\bm{I})$ to $\bm{x}_{T}$, where $T$ is the number of sampling time steps.

\section{tLP and tOF}
In \cite{tOFtLP}, pixel-level video flows tOF and perceptual video flows tLP are respectively defined as:
\begin{align}
    \text{tOF} &= \sum_{\text{i=1}}^{\text{L}} ||OF(\hat{g}_{{i-1}}, \hat{g}_{{i}})-OF(g_{{i-1}}, g_{{i}})||_1, \\
    \text{tLP} &= \sum_{\text{i=1}}^{\text{L}} ||LPIPS(\hat{g}_{{i-1}}, \hat{g}_{{i}})-LPIPS(g_{{i-1}}, g_{{i}})||_1,
\end{align}
where $g_\text{i}$ is \textit{i}-th frame of ground truth video, $\hat{g_\text{i}}$ is \textit{i}-th frame of restored video, $OF(\cdot)$ is optical flow estimator and $LPIPS(\cdot)$ is LPIPS score estimator. We use the RAFT \cite{raft} model to estimate $OF(\cdot)$. 

\begin{table}[t]
  \caption{
  Comparisons with previous guidance methods using Inception Score (IS)~\cite{is} and Fr\'echet Inception Distance (FID)~\cite{fid} metrics. feed-forward/iter. indicates the number of feed-forward passes required by the diffusion model for each denoising iteration. The best and second-best scores are marked in\textbf{bold} and \underline{underline}.}
  \label{tab:image_synthesis}
  \begin{tabular}{c|ccc}
    \Xhline{2\arrayrulewidth}
        Metrics & {SAG~\shortcite{sag}} & {PAG~\shortcite{pag}} & {\drgabb} \\
        \hline\hline
        IS $\uparrow$ & 33.70 & \textbf{36.76} & \underline{36.04} \\
        FID $\downarrow$ & 16.31 & \underline{16.27} & \textbf{15.20} \\
        \hline
        feed-forward/iter. $\downarrow$ & \underline{3} & \underline{3} & \textbf{2} \\
    \Xhline{2\arrayrulewidth}
  \end{tabular}
\end{table}

\begin{table}[t]
  \caption{
  Diffusion inference time (in seconds/frame) comparison with previous diffusion-based video super-resolution tasks on $4\times$ upscaling 640$\times$360 video.
  }
  \label{tab:inferencetime}
  \begin{tabular}{c|ccc}
    \Xhline{2\arrayrulewidth}
        Methods & UAVideo \shortcite{zhou2024upscaleavideo} & MGLD \shortcite{yang2023mgldvsr} & Ours \\
        \hline\hline
        Time (sec./frame) & {31} & {54} & {67} \\
    \Xhline{2\arrayrulewidth}
  \end{tabular}
\end{table}

\end{document}